# Towards Integrating True Random Number Generation in Coherent Optical Transceivers

Dinka Milovančev[(1)], Nemanja Vokić[(1)], Christoph Pacher[(1)], Imran Khan[(2)], Christoph Marquardt[(2)], Winfried Boxleitner[(1)], Hannes Hübel[(1)], and Bernhard Schrenk[(1)]

[(1)] *AIT Austrian Institute of Technology, Center for Digital Safety&Security, Giefinggasse 4, 1210 Vienna, Austria. Author e-mail address: bernhard.schrenk@ait.ac.at*

[(2)] *Max Planck Institute for the Science of Light, Quantum Information Processing Group, Staudtstr. 2, 91058 Erlangen, Germany*

The integration of quantum communication functions often requires dedicated opto-electronic components that do not bode well with the technology roadmaps of telecom systems. We investigate the capability of commercial coherent transceiver sub-systems to support quantum random number generation next to classical data transmission, and demonstrate how the quantum entropy source based on vacuum fluctuations can be potentially converted into a true random number generator for this purpose. We discuss two possible implementations, building on a receiver- and a transmitter-centric architecture. In the first scheme, balanced homodyne broadband detection in a coherent intradyne receiver is exploited to measure the vacuum state at the input of a 90-degree hybrid. In our proof-of-principle demonstration, a clearance of >2 dB between optical and electrical noise is obtained over a wide bandwidth of more than 11 GHz. In the second scheme, we propose and evaluate the re-use of monitoring photodiodes of a polarization-multiplexed inphase/quadrature modulator for the same purpose. Time-interleaved random number generation is demonstrated for 10 Gbaud polarization-multiplexed quadrature phase shift keyed data transmission. The availability of detailed models will allow to calculate the extractable entropy and we accordingly show randomness extraction for our two proof-of-principle experiments, employing a two-universal strong extractor.

## 1. Introduction

The security of many applications in the field of information and communication technology, such as data encryption or digital signatures, strongly depends on random numbers that are used as seed for their primitives. This implies that the highest quality of randomness is needed in order to guarantee digital security. One of the vital tools to provide such true randomness is quantum random number generation (QRNG) [1-3]. QRNG builds on the fact that quantum states provide, if measured correctly, inherent randomness. In particular, optical implementations of QRNGs exploit the quantum properties of light in order to provide randomness from intrinsic quantum mechanical features that are connected to Heisenberg's uncertainty relation.

This method of generating unpredictable random numbers, which are sourced by means of photonics, requires dedicated opto-electronic signal converters. Several approaches have been investigated in earlier works. On the one hand, single-photon avalanche photodetectors (SPAD) can be used to acquire the path choice of a single photon [4], to measure the detection time interval of photons [5,6], or both [7]. Alternatively, the photon number or the spatial distribution can be resolved [8-11]. On the other hand, balanced homodyne detectors can be exploited to measure the vacuum fluctuation of an optical field [12-14], the laser phase noise [15] or the randomized phase of switched laser pulses [16].

The efficiency of these schemes strongly relies on the performance of the employed detector. SPAD devices are typically operated at detection rates equal or below 1 GHz [17,18]. Furthermore, SPADs are not integrated well with mainstream telecom technology development. This is further hindered by the missing technological overlap with respect to integration platforms and operation at low temperatures, which as techno-economic roadblock is even more pronounced for the faster super-conducting single-photon detectors. Therefore, the practical implementation of SPADs to obtain an integrated QRNG functionality in existing communication systems is rendered as cost-inefficient.

Balanced homodyne detectors are widely adopted in the field of coherent optical communications and have been laid out for high opto-electronic bandwidths that are already exceeding 10 GHz and are available as commercial products. However, in order to be applicable to the quantum domain, the receiver has to comply with the requirements in terms of noise, in particular concerning the transimpedance amplifier (TIA) that is used as signal converter for the detected photocurrent. Most previous implementations of QRNGs have therefore relied on custom detector implementations with a rather low bandwidth of up to 1 GHz [19,20], without resorting to commercially deployable receivers.

In this work we exploit such a broadband coherent receiver as multi-purpose element. Next to classical signal reception, it enables the acquisition of quantum noise (of the I and/or Q component of the electro-magnetic field) through use of the local oscillator (LO), yielding an optical-to-electrical noise clearance of >2 dB over a wide bandwidth of 11.8 GHz. To complete the experiment, we have implemented a strong randomness extractor to experimentally extract random bits from the measurement data using an estimate of the extractable length.

We further demonstrate that QRNG functionality can also be accomplished in an optical inphase/quadrature (I/Q) modulator, as it is commonly adopted in coherent transmitters. Building on a similar methodology as for the coherent receiver arrangement, we demonstrate random number extraction from homodyne measurements of the quantum vacuum state, time-interleaved with classical transmission of quadrature phase shift keyed (QPSK) data and run a random number test suite to rule out obvious problems in our random number extraction implementation.

The contribution of this paper is to demonstrate that QRNG functionality can be integrated in state-of-play opto-electronic transmitter and receiver components, which have not been intentionally laid out to support quantum-optic applications. In this way, deployable opto-electronic technology, which has been proven in the field under the umbrella of coherent optical telecommunication networks, are exploited as multi-purpose element that requires just a marginal cost spending to practically implement QRNG in classical transceivers.

The paper is organized as follows. Section II introduces the methodology to generate random numbers building on multi-purpose photonic components as found in coherent communication links. Section III details the experimental setting to exploit a coherent receiver as QRNG, while Section IV presents the experimental results. Section V then integrates QRNG functionality in an optical transmitter, which is naturally not laid out for signal reception. Section VI presents the experimental arrangement for this purpose and Section VII discusses the results. Classical data transmission in virtue of multi-purpose transceiver operation is demonstrated in Section VIII, using the same sub-systems for transmitter and receiver as employed for the two QRNG schemes. Finally, Section IX concludes the work.

## 2. QRNG Integration for Coherent Receivers

QRNG can serve as a seed engine for various communication primitives. For this purpose, both parties of the optical communication link may employ QRNG functionality integrated in either the receiver or the transmitter (Fig. 1). Coherent receivers can be conveniently re-used for the purpose of QRNG. Such a QRNG builds on the measurement of the vacuum state of a blanked signal input in a homodyne detector, i.e., at the signal port of the optical mixer that is used to beat the LO with a potentially present signal [12] or on hybrid detection of an arbitrary quantum state [21]. Given a model that allows to calculate a lower bound for the (conditional min-) entropy, such quantum entropy sources can be converted into a true random number generator.

Broadband operation in the range of 10 GHz can be in principle supported through this method [22], provided that the electrical noise contribution of the corresponding TIA technology allows for a sufficient quantum-to-classical noise clearance, as sketched in the inset of Fig. 1 for the case of the electrical background noise and an additional optical contribution corresponding to the noise of the seed light. Balanced operation for both detector branches in both, optical power and skew, enables TIA operation with a high transimpedance gain without saturation effects for the detector. Moreover, it suppresses crosstalk noise from side-channels that can potentially co-exist out-of-band to the optical detection window that is defined by the LO [23]. Although the latter situation does not apply to implementations that exclusively aim at QRNG functionality, this aspect has to be taken into consideration when multi-purpose operation of a classical coherent receiver is sought for.

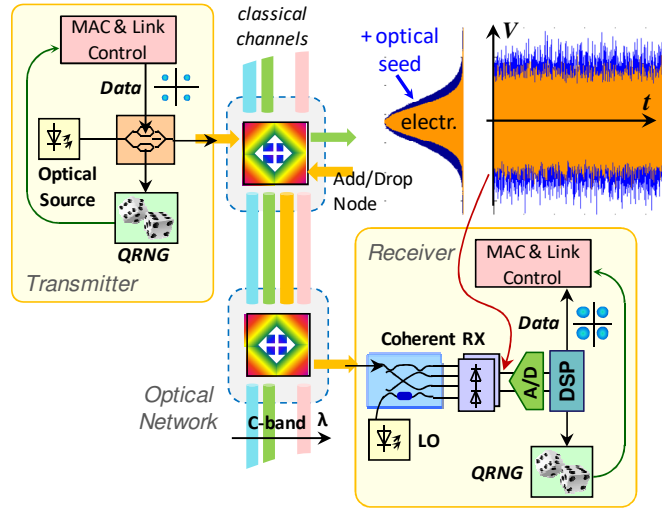

Fig. 1. Coherent communication link with integrated randomness engines.

## 3. Experimental Setup

The QRNG function of a coherent receiver as used in telecom systems is evaluated in the coherent link shown in Fig. 2(a). At the receiving site, coherent intradyne detection is performed with a polarization-multiplexed (PM) 90-degree hybrid as optical mixer and two balanced receivers (BRX, model Discovery Semiconductors, DSC-R405) for each polarization tributary, yielding the respective I and Q quadratures after digital signal processing (DSP). Optical channel selection is made through the inherent filtering of the coherent detection process. For this purpose, the LO is tuned to the optical carrier frequency of the target channel within the C-band.

The coherent optical link was loaded with 11 PM-QPSK channels, following the dense wavelength division multiplexed (DWDM) grid in the spectral range from 1548.51 to 1560.61 nm (see inset in Fig. 2(a)), with a minimum spacing of 93.8 GHz. The comb of channels was modulated with a PM-I/Q modulator using an arbitrary waveform generator (AWG). The QPSK symbol rate was 10 Gbaud, meaning a capacity of 40 Gb/s/λ. The comb was boosted by an Erbium-doped fiber amplifier (EDFA) before being launched with 2 dBm/λ to the link span. A variable attenuator ($A_{tt}$) before the receiver allows to investigate loading effects for later QRNG operation.

The QRNG function demonstrated at the receiver site is based on the multi-purpose operation of the coherent receiver. An LO with an optical power of 18 dBm is exploited as the seed light source for the proof-of-principle QRNG, which further builds on the balanced detectors to acquire the shot-noise of the LO. In the case of QRNG operation, only the LO input is required. If only one balanced detector is to be used, any input signal should be strictly avoided since an adversary could take control of it and potentially spoil the randomness, e.g., by using non-classical squeezed light with reduced quantum noise in the quadrature that is being measured. However, at this moment it shall be assumed that non-classical light sources are not available for an adversary. Shortly, we will discuss how we can get rid of this assumption. We spectrally parked the LO at 1528 nm for this adversary-free setting, far off from the active DWDM channels and out of the EDFA passband range so that amplified spontaneous emission resulting from the optical amplification is filtered. With this setting, any present side-channel is principally opto-electronically filtered. In addition, the side-channel noise due to direct-detection terms is suppressed by ensuring a high-enough common-mode rejection ratio (CMRR) for the balanced detectors.

The employed balanced detectors are rated for 10 Gbaud data transmission. The detectors have a bandwidth of 14.8 GHz for each of the two photodiodes, a responsivity of 0.8 A/W and a transimpedance gain of 51.5 dBΩ. Although this high bandwidth is not required to perform QRNG functionality, it is a prerequisite for modern coherent communication systems. The wide bandwidth also necessitates the co-integration of the balanced detector with a broadband TIA, which is not noise-optimized for the QRNG application.

Figure 2(b) shows the RF response at the detector output under common-mode injection. The imbalance corresponds to the case where one of the photodiodes is disconnected. In contrast, balancing of the delivered optical power to both photodiodes leads to a suppression in the electrical signal output in virtue of the common-mode rejection of the balanced detector. The corresponding CMRR reaches a value of up to 34 dB at 6 GHz. In order to investigate an imperfect coherent receiver, the CMRR has been intentionally degraded, down to a worst-case setting of 9.5 dB.

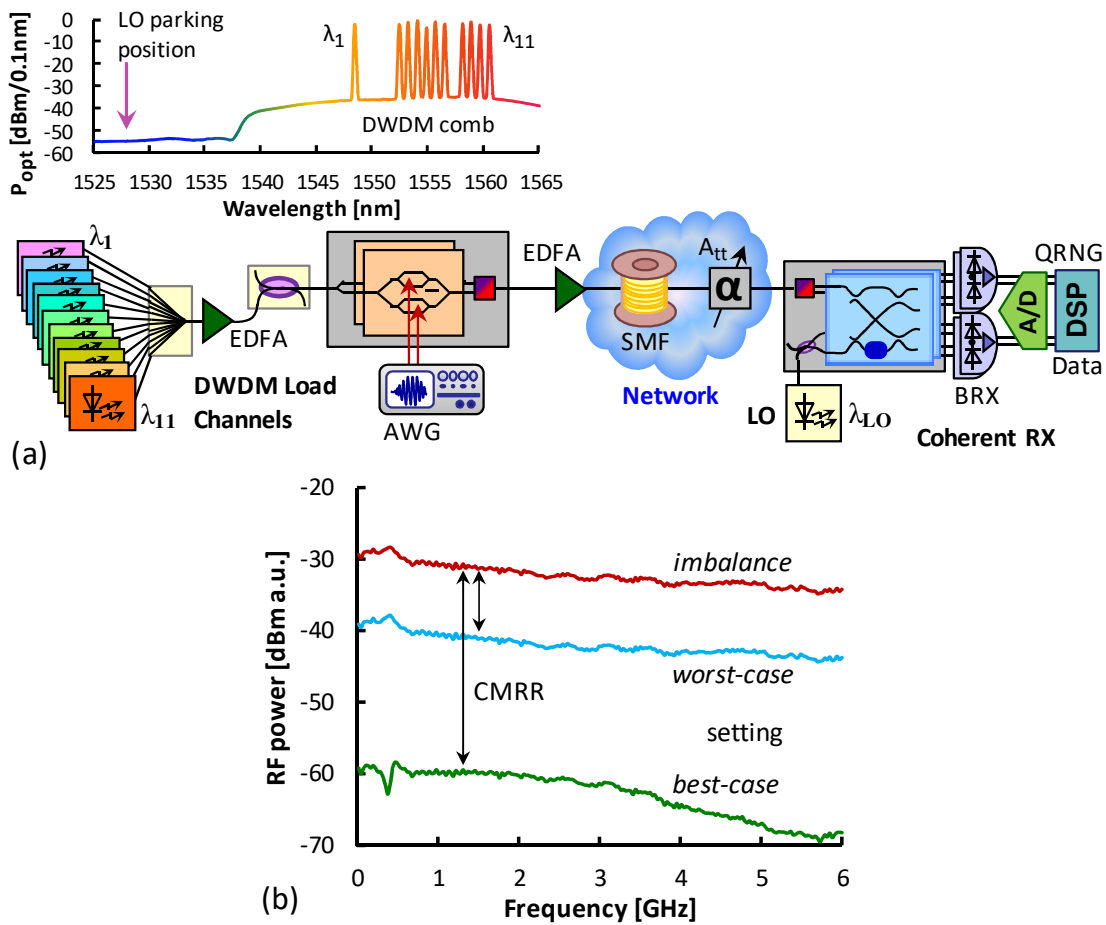

Fig. 2. (a) Experimental setup for receiver-side QRNG.
(b) Response of a balanced detector employed at the coherent receiver.

The balanced detector outputs are then converted from the analogue to the digital (A/D) domain in course of the acquisition through a real-time scope. The sampled bits are, however, not uniformly random. The reasons for that are manifold: (i) the samples are Gaussian distributed, (ii) the non-ideal transmission function of the TIA or any post-amplifier leads to correlations between samples, (iii) the amplifiers could have some deterministic behavior that leads to a hidden pattern in the electronic noise, (iv) any form of crosstalk that may result from optical or electrical coupling. In order to enhance the quality of the random numbers of our experiment, the sampled data has been the input to a seeded (strong) randomness extraction algorithm [24]. This algorithm improves the entropy per bit by hashing the random data using an independent random seed. Since we used a strong extractor that seed can be re-used in subsequent extraction processes. Our proof-of-principle randomness extraction process has been performed in terms of off-line DSP using a rough estimate of 1/8 bit per sample of the min-entropy of the data, which is equal to the negative log of the guessing probability [25].

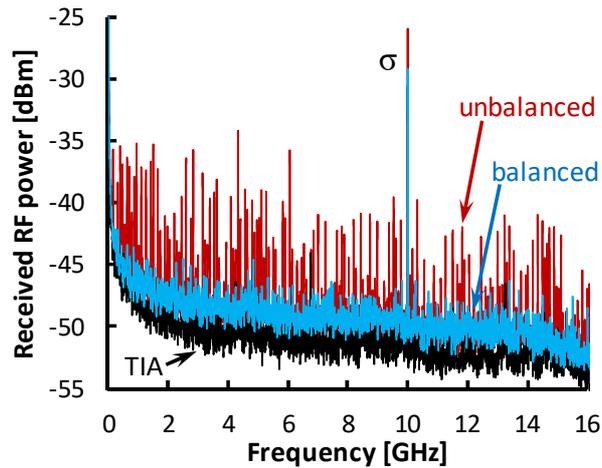

Fig. 3. Received RF signal spectrum at balanced detector output without and with optical input, and for degraded CMRR.

## 4. QRNG Performance of Coherent Receiver

The generation of random numbers has been first investigated in a back-to-back scenario in which the optical link is simply emulated by optical attenuation. The impact on a CMRR degradation has been included in order to evaluate the impact of side-channel crosstalk, as it appears in a loaded coherent DWDM link. Figure 3 presents the received RF spectrum of a balanced detector and reports the effect of a reduced CMRR. Without optical input, the noise of the TIA determines the electrical background level. For the perfectly balanced receiver with high CMRR and a high delivered average optical power of -15.4 dBm/λ for the modulated side-channels, the noise increases but remains without spurious spectral components. Compared to the considerably lower reception sensitivity for 10 Gbaud PM-QPSK transmission, which will be discussed in Section VIII, this proves that a high dynamic range of >13 dB is compatible and would allow to add more classical channels. The quantum-to-classical contrast in terms of power level between optically lit and dark receiver, which is >2 dB over a bandwidth of 11.8 GHz, corresponds to the earlier introduced clearance. In the worst case of a partially unbalanced detector with a low CMRR of 9.5 dB, data harmonics of the side-channels are clearly visible above the shot-noise level. It shall be stressed that the strong harmonic at 10 GHz (σ) derives from the sampling of the signal at a quadrupled rate of 40 GSa/s.

These artifacts in the acquired signal require randomness extraction in order to obtain truly random numbers. For this purpose, 8M samples have been recorded at 8-bit resolution. Universal Toeplitz hashing has been employed as seeded randomness extractor [26]. As a first step we estimated the lower bound on min-entropy to be 1/8 bit per acquired sample. In a deployed system the estimation of entropy has to be derived from a full physical model of the optical and electrical components. A block of 1 Mibit = 1,048,576 bits from 64 Mibit of sampled data has been extracted. Based on the aforementioned 1/8 bit per sample between raw and extracted bit rate, we estimate that we could generate approximately $2 \times 10^7$ 256-bit keys per second and detector. Given that four balanced detectors are employed in a coherent intradyne receiver, this rate can be further elevated. An exemplary 256-bit random number string is shown in Table I.

As a sanity check for the extraction process we run the complete NIST SP800-22-rev1a randomness test suite [27] before randomness extraction and after extraction. The number of failed NIST tests is ~39 out of 188 without extraction, including the Block Frequency, the Runs, the Longest Run of Ones, the Binary Matrix Rank and the Discrete Fourier Transform tests, 29 of the Non-overlapping Template Matching tests, the Overlapping Template Matching, the Approximate Entropy, one Random Excursions, the first and second Serial tests. This high number reduces to ~3 when the extraction is employed, including two Non-overlapping Template Matching and one Random Excursions test.

In order to further evaluate the impact of Raman scattering from the remote DWDM channels in the upper C-band above 1548 nm into the LO reception band at 1528 nm, the optical link loss has been partially eroded by 40.3 km of standard single-mode fiber. Similarly as for the aforementioned back-to-back scenario, random numbers can be generated, with the extraction process check resulting in identical results.

If a complete suppression of the input signal to the receiver is not possible and if the input signal cannot be trusted, as initially discussed before, it is required to measure both, I and Q components, simultaneously, in order to be able to leverage the quantum noise of the unknown input signal [21].

| QRNG | Sample 256-bit random bit sequence | | | |
|---|---|---|---|---|
| Coherent RX | 00110010 | 00110000 | 10110111 | 01011010 |
| | 00110001 | 11110010 | 00001001 | 00010110 |
| | 11010011 | 10111101 | 00011111 | 00011010 |
| | 01000111 | 01111100 | 11111110 | 01011001 |
| | 00110110 | 00101011 | 00001000 | 01010000 |
| | 00111000 | 00101111 | 11100011 | 11100100 |
| | 10110110 | 11000000 | 10001100 | 11100010 |
| | 11100110 | 00111001 | 10100111 | 01100111 |
| PM - I/Q Modulator | 00001101 | 01011000 | 00001001 | 01011000 |
| | 10001001 | 10000001 | 01010011 | 11100111 |
| | 11011001 | 11011011 | 11001011 | 01011001 |
| | 00010110 | 10100011 | 10010010 | 10111010 |
| | 10010010 | 10110111 | 10110101 | 11010101 |
| | 00000011 | 01001100 | 00100011 | 11100010 |
| | 11101100 | 11011100 | 01101010 | 10000101 |
| | 10011010 | 01101101 | 10011110 | 01010111 |

Table I. Generated Random Bit Sequences

## 5. QRNG Overlay in a Polarization-Multiplexed I/Q Modulator

In contrast to coherent receivers, transmitting sub-systems are not laid out to perform signal reception, and particularly not at the quantum level. Nevertheless, we prove the concept of balanced homodyne detection with a commonly adopted optical transmitter by exploiting the integrated signal monitors in an optical I/Q modulator.

The OIF implementation agreement [28] foresees a PM-I/Q modulator to tap and monitor the two optical outputs of its nested Mach-Zehnder modulators (MZM). The integrated photodetectors are typically featuring an opto-eletronic bandwidth in the range of 0.5 to 1 GHz for this purpose. Based on the assumed ratio of 1/8 bit per sample, this bandwidth would potentially support, together with a 10% tap ratio, a QRNG rate of up to 500 Mb/s.

Figure 4(a) presents the QRNG overlay in such a PM-I/Q modulator (model Sumitomo T.SBXH1.5PL-25PD-ADC). The parental configuration of the modulator is solely used for the purpose of power splitting its source laser input (A) to each of the nested child-MZMs of each polarization branch. The latter serve as tunable attenuator (B) in virtue of their interferometric setup by electrical tuning of the I,Q bias points of each child. With this, parent and child form a high-precision 50/50 splitter as it is required for a balanced detector arrangement. The two monitor photodetectors (C) at the polarization branches complete this balanced homodyne detector (D). The balanced homodyne detector that is formed by the I/Q modulator elements is complemented by a 66-dBΩ TIA with differential feedback loop that is closed around an 8-GHz gain-bandwidth operational amplifier. It uses a balanced configuration at its electrical input, which is fed by the two single-ended PIN photodiodes.

Compared to a coherent receiver as applied in Sections III-IV, the accomplishment of a high clearance between optical and electrical noise is now challenged by the low tap ratio of the monitor photodetectors and the modulator losses. Similar as for the receiver-integrated QRNG function, time division multiplexing (TDM) would have to be applied to classical PM-QPSK transmission in order to periodically generate random numbers through use of a multi-purpose photonic transmitter. This will be demonstrated in Sections VI and VII.

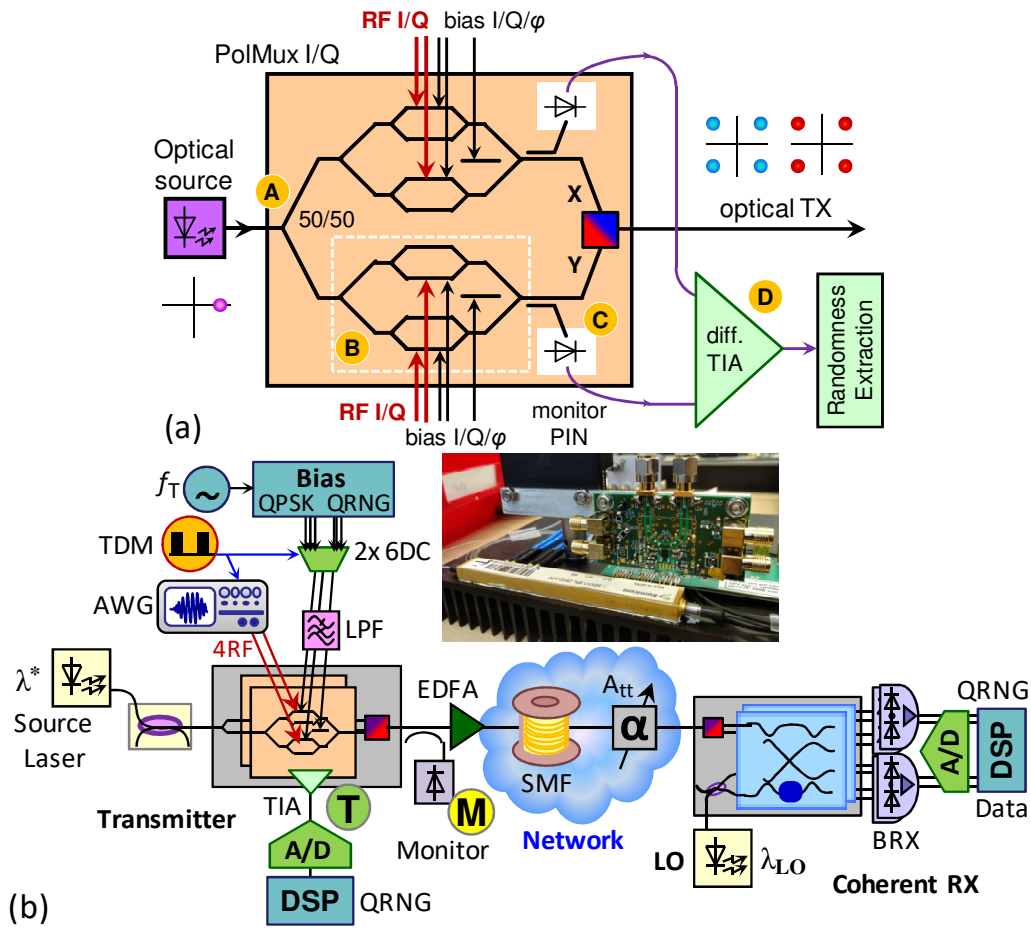

Fig. 4. (a) Random number generator overlay for a PM-I/Q modulator. (b) Experimental setup for joint QPSK transmission and transmitter-side QRNG. The inset shows the PM-I/Q modulator and the TIA that has been interfaced with the monitor photodiodes of the modulator.

## 6. Experimental Characterization of the Transmitter-Integrated QRNG Engine

The clearance that can be obtained for a given source laser input is determined by the noise of the dark I/Q modulator in QRNG configuration. Figure 5(a) presents the electrical noise spectrum ($\varepsilon$) of the TIA when the source laser is not lit. In order to surpass this background noise level when the optical source of the I/Q modulator is switched on, the bias point of the modulator needs to be optimized. Each of the child MZMs is operated so that the photocurrent to the monitor photodiodes is maximized and a high TIA output is yielded. Moreover, the relative adjustment of the child MZM outputs ensures balancing, thus avoiding saturation of the TIA.

As a result, the acquired noise surpasses the electrical TIA background for an optical source laser level of 14 dBm or higher, over a bandwidth of ~150 MHz. The dependence of the clearance on the source power level is reported in Fig. 5(b). We obtained a clearance of 2.1 dB at 18 dBm. A linear dependence between clearance and source power indicates that the generated noise derives from the vacuum fluctuations. The combination of high source power level and high transimpedance requires a careful adjustment of the modulator bias points. The admissible range for the I,Q bias points of the child MZMs was typically 0.05 $V_\pi$ at the optimal setting, in order to avoid imbalance and saturation of the TIA.

Moreover, a few electro-magnetic interference notes can be observed in the acquired noise spectrum. After opto-electronic detection and signal conditioning a randomness extraction is therefore performed, similarly as it was included for the receiver-centric QRNG.

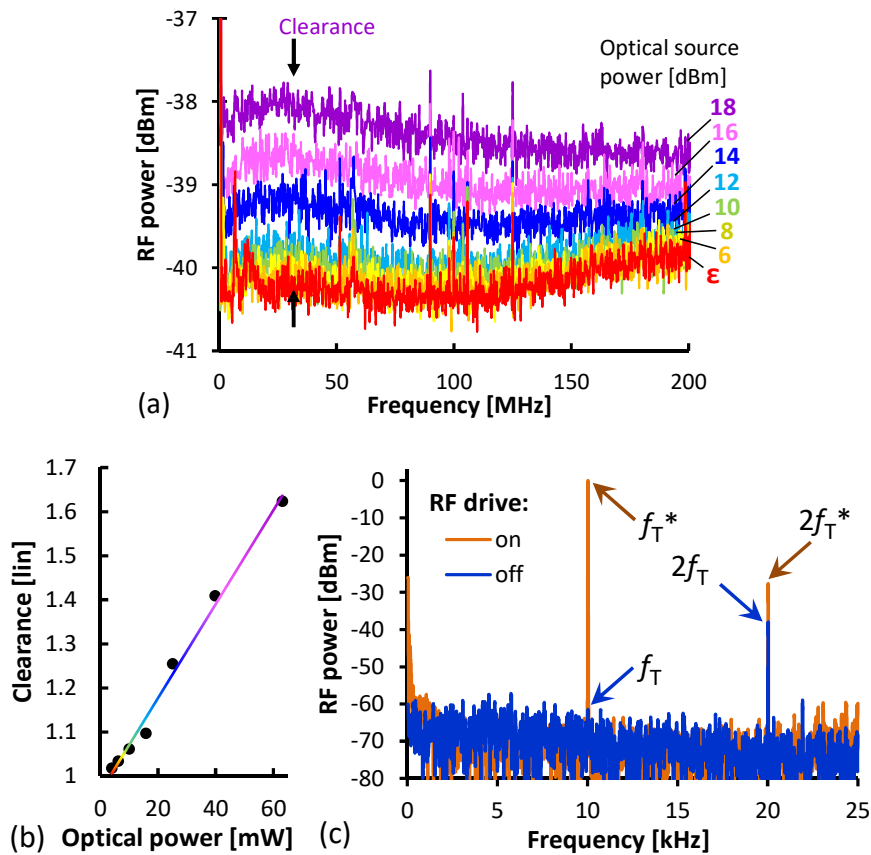

Fig. 5. (a) Received RF spectra due to quantum noise and electrical ($\varepsilon$) TIA background, and (b) corresponding clearance. (c) Bias tone spectrum.

## 7. Time-Interleaved QRNG Function in Optical I/Q Transmitter

The experimental setup in Fig. 4(b) has been used to evaluate random number generation time-interleaved with 10-Gbaud PM-QPSK data transmission. The practical investigation of time sharing opto-electronic components for two very different use cases, namely coherent optical broadband communications and QRNG, is important as it assesses the compatibility of state-of-play component technology to accommodate both rather than just one of the operation modes. As will be discussed in this section, switching transients and the opto-electronic characteristics require a careful design of a time-shared I/Q transmitter that is exploited as multi-purpose hardware for both, PM-QPSK transmission and QRNG.

In order to account for the bias point optimization at the transmitter-centric QRNG, two sets of time-multiplexed bias points are applied to the I/Q modulator. These sets are dedicated to PM-QPSK and QRNG operation, respectively, and are optimized for the actual mode of the modulator – with and without applying a radio frequency (RF) drive signal. The reason to do so is the insufficient electrical isolation between RF and DC electrodes of the modulator, which leads to a crosstalk of -7.6 dB. According to this high crosstalk, we observed a shift in the DC bias point when switching on the RF drive at the adjacent electrode. This effect is seen in Fig. 5(c), which presents the spectrum of a DC bias tone that is modulated at the DC electrode with a frequency of $f_T$ = 10 kHz. Operation in the null point of the transfer function of the modulator, without being driven by an RF signal, results in a minimized fundamental tone $f_T$ and a noticeable second harmonic at $2f_T$. When the RF drive is switched on, the fundamental tone is again pronounced ($f_T^*$), corresponding to an average shift in the bias point of ~1.56V or $0.42V_\pi$. This shift needs to be accounted for due to the TDM operation for PM-QPSK for which the QPSK drive is blanked during QRNG operation for which an opto-electronic bandwidth within the QPSK signal spectrum is allocated.

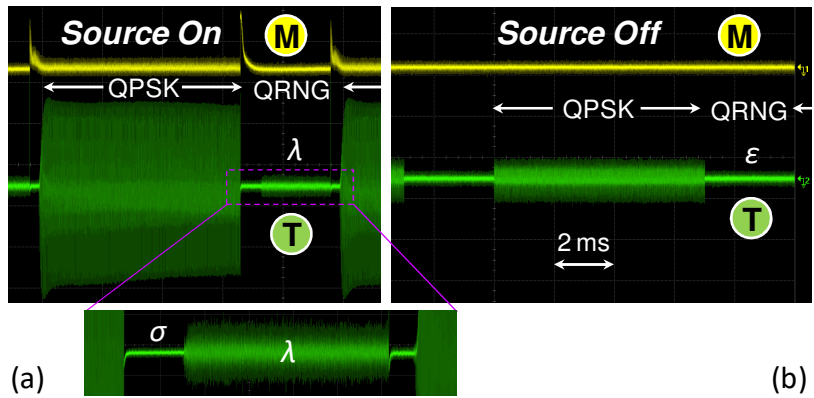

Fig. 6. Acquired TDM frame with (a) and without (b) optical seed.

A TDM frame with a period of 10 ms and a duty cycle of 70% for data transmission has been applied at the transmitter. This TDM timing applies for example when multiple transmitters in an optical network with multipoint-to-multipoint architecture share a coherent receiver. At the same time, it leaves a 3-ms window to generate random numbers in every frame.

Figure 6(a) shows the signal traces acquired through an optical intensity monitor (M) at the output of the PM-I/Q modulator, and at the electrical TIA output (T). During data transmission, the QPSK modulation is clearly visible at the TIA output. The overlap of data and quantum noise spectrum prevents QRNG functionality. Therefore, the QPSK data is blanked during the QRNG slot of the TDM frame. The quantum noise signature ($\lambda$) is then clearly visible and can be acquired for random number generation. At the transitions between the two TDM slots, the bias points of the modulator are switched, which temporarily saturates the TIA due to the imbalance in the homodyne detector configuration until the bias point settles. This effect is observed as short null-line in the time trace ($\sigma$).

Figure 6(b) presents the signal traces for deactivated optical source laser. Instead of a well-pronounced optical noise contribution (see λ in Fig. 6(a)) according to the clearance reported earlier, the noise at the TIA output reduces to the electrical background noise level (ε). The aforementioned electrical crosstalk between RF and DC electrodes of the modulator leads to a residual QPSK modulation during the data-oriented TDM slot.

The transients that appear at the edges of the TDM channels (i.e., time slots) are induced by lowpass filters (LPF) at the DC bias lines of the modulator, which had a 3-dB bandwidth of 22 kHz. These LPF have been inserted in order to suppress electro-magnetic interference and thus to prevent electrical excess noise at the QRNG output. However, although the bandwidth allows for bias tone probing, it limits the native 700-kHz electro-optic response of the DC electrodes, which leads to transients in the signal trace when switching between two sets of bias points.

A total of 5.4M samples have been recorded during TDM operation in QRNG mode. Randomness extraction at the aforementioned estimated lower bound of 1/8 bit per acquired raw sample has been conducted on a block of 1 Mibit (4096 keys of 256 bit each). The NIST test suite has been applied twice: before randomness extraction 69/188 tests did not pass. After extraction, all but one test passed successfully, confirming the sanity of the randomness extraction step. Based on the aforementioned ratio of 1/8 between raw samples and extracted bits, we estimate that for the given TDM duty cycle $\sim 9 \times 10^4$ 256-bit keys can be generated each second. An exemplary 256-bit random number string obtained through the transmitter-integrated QRNG is appended to Table I.

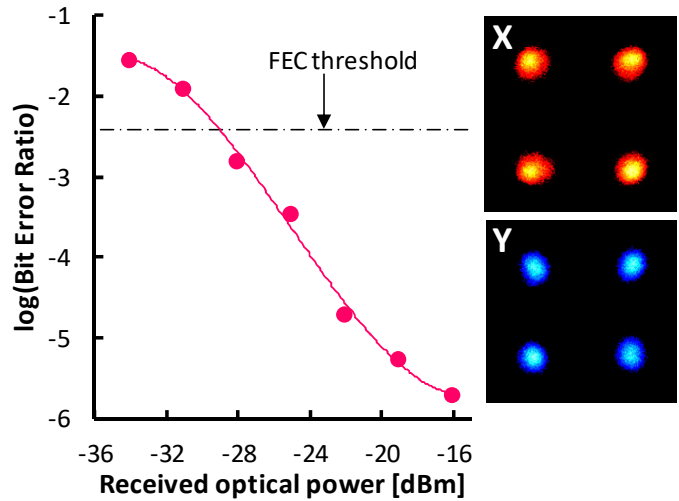

Fig. 7. BER performance for coherent transmission link and received PM-QPSK constellation.

## 7. PM-QPSK Transmission

For the sake of completeness, classical data transmission has been evaluated for the coherent link in virtue of the multi-purpose transmitter and receiver architecture. 10 Gbaud PM-QPSK has been transmitted together with the load channels over the 40.3 km fiber-based link and received following a coherent intradyne reception methodology with off-line frequency offset compensation, carrier-phase recovery and decision sampling. The transmission performance has been evaluated for an optical source wavelength of 1549 nm, which is situated within the spectral range of the DWDM comb. The optical signal-to-noise ratio was 37.2 dB and the bit error ratio (BER) has been evaluated as function of the received optical power.

Figure 7 presents the BER performance for this 40 Gb/s link. The sensitivity at the hard-decision forward error correction (FEC) threshold of $3.8 \times 10^{-3}$ is -29.1 dBm. The constellation diagrams, which are appended to Fig. 7, show a clean QPSK constellation in both polarization planes. This proves the multi-purpose functionality of PM-I/Q modulator and coherent receiver for data transmission and for QRNG.

## 8. Conclusion

We have proposed the multi-purpose use of coherent transceiver sub-systems for data transmission and quantum random number generation. Taking advantage of broadband balanced homodyne detectors in a coherent intradyne receiver, the vacuum state has been measured in virtue of the high LO power level that is supported in state-of-play coherent receivers. Given the obtained optical-to-electrical noise clearance of >2 dB over a wide bandwidth of 11.8 GHz, this receiver-centric QRNG promises high generation rates in combination with a Toeplitz randomness extractor.

On top of this, we have experimentally demonstrated the re-use of an optical PM-I/Q modulator for the purpose of QRNG. By leveraging its integrated monitor photodiodes, the vacuum fluctuations of the unused input state have been acquired through use of the transmitter-side source laser. TDM operation with a PM-QPSK data stream with time-interleaved QRNG has been validated.
Future work may include full information-theoretic models of the proposed QRNGs, in order to obtain certifiable random number generators.

## 9. Acknowledgement

This work was supported in part by funding from the European Union's Horizon 2020 research and innovation programme under grant agreement No 820474 and the European Research Council (ERC) under grant agreement No 804769.